\documentclass[%
 reprint,
 amsmath,amssymb,
 aps,
]{revtex4-2}

\usepackage{graphicx}
\usepackage{dcolumn}
\usepackage{bm}

\usepackage{xcolor}
\usepackage{soul}

\usepackage{color,soul}
\usepackage[percent]{overpic} 
\usepackage{threeparttable}

\usepackage{longtable}
\usepackage[autostyle]{csquotes}
\usepackage{array,booktabs} 
\usepackage{mathtools}
\usepackage{bm}

\begin{document}

\preprint{APS/123-QED}

\title{Thermodynamic Origin of Water's Thermal Conductivity Maximum}

\author{Oliver R. Gittus}
 \email{o.gittus18@imperial.ac.uk}

\author{Fernando Bresme}%
 \email{f.bresme@imperial.ac.uk}
\affiliation{Department of Chemistry, Molecular Sciences Research Hub, Imperial College London, London W12 0BZ, United Kingdom.}%


\begin{abstract}
The thermal conductivity of water features a maximum (TCM) as a function of temperature at constant pressure.
By examining why molecular force fields succeed or fail to reproduce the maximum and interpreting our results using the Bridgman equation, we show that water's TCM is connected with its compressibility minimum. Using Stillinger-Weber potentials for tetrahedral liquids, we interpolate between the behaviour of simple liquids and highly tetrahedral materials such as carbon. Together with two vanishing limits at low/high tetrahedrality, we identify three regimes for the TCM: when it originates from either the compressibility minimum or density maximum, or both. Thus, the TCM exists in a \enquote{Goldilocks Zone} of tetrahedral order. We provide a thermodynamic explanation for the TCM of not only water, but tetrahedral liquids in general.

\end{abstract}

\maketitle


At constant pressure the thermal conductivity (TC) of liquid water increases with temperature until it reaches a maximum at {{404}~K} (10~bar)~\cite{Huber2012_WaterTC}, then decreases upon further heating until the boiling point is reached. In contrast, the TC of a simple liquid monotonically decreases with increasing temperature primarily due to the corresponding decrease in density. 
Thus, water's thermal conductivity maximum (TCM) and subsequent decrease upon cooling are anomalous properties.   
The temperature of the TCM increases with pressure; all natural and industrial processes that occur at near-standard conditions therefore correspond to the anomalous heat transport regime of water. 
Despite its significance, the physical origin of water’s TCM remains an open question.

Theoretically predicting the TC of liquids from first-principles is a challenging problem. In arguably the earliest attempt (1923), Bridgman connected the TC of a liquid to its isentropic speed of sound.
He imagined liquid molecules arranged in a cubic lattice, with the internal energy difference due to the temperature gradient being \enquote{handed down a row of molecules at a rate determined by the speed of sound}~\cite{Bridgman1923_TCeq}. This was the first in a family of quasi-lattice models that assume liquid molecules oscillate about fixed points in a solid-like lattice on the time scale of heat transport, and exchange energy via nearest-neighbour collisions~\cite{McLaughlin1964_TC,Chen2020_ThermophysicsLiquids,Xi2020_TCmodel,Zhao2021_liquidTC,Khrapak2021_liquidTC}. When the characteristic frequency of energy exchange is identified with the acoustic spectrum, these models give a TC of the form $\lambda_\mathrm{QL} \propto \delta^{-2} c $, where $\delta$ is the average distance between molecules and $c$ is the speed of sound~\cite{Khrapak2021_liquidTC,Zhao2021_liquidTC}. 
While these models may give accurate predictions for specific liquids at specific thermodynamic conditions, they cannot in general predict the TC of liquids~\cite{McLaughlin1964_TC,Xi2020_TCmodel,Zhao2021_liquidTC,Khrapak2023_BridgmanTest,Zhang2023_waterTCextreme}.
However, we will show that when used together with simulations, they provide insight into the origins of water's TCM. 

We consider one of simplest quasi-lattice models: the Bridgman equation empirically corrected for polyatomic molecules, $\lambda_\mathrm{B} = 2.8 k_\mathrm{B} \delta^{-2} c_S$, which using the thermodynamic relations $c_S^2 = (\rho \beta_S)^{-1}$ and {$\beta_T / \beta_S = C_P/C_V = \gamma$}, can be written as
\begin{equation}\label{eq:bridgman}
    \lambda_\mathrm{B} = 2.8 k_\mathrm{B} \left(\frac{\rho \gamma^3}{M^4 \beta_T^3}\right)^{1/6}
\end{equation}
where $\rho$, $M$, $\beta_T$ ($\beta_S$) and $\gamma$ are the density, molecular mass, isothermal (isentropic) compressibility and adiabatic index, respectively. {$C_P$ and $C_V$ are the isobaric and isochoric heat capacities, respectively}.
$\rho(T)$ and $\gamma(T)$ both increase monotonically in the temperature range that the extrema in $\lambda_\mathrm{B}$, $c_S$, $\beta_S$ and $\beta_T$ occur. Thus, all four extrema share the same phenomenological origin, i.e., if one did not exist, then they all would not exist, and the effect of $\rho(T)$ and $\gamma(T)$ is to shift the temperature $T_\mathrm{ex}$ of the extrema. Thus, the Bridgman equation, along with other quasi-lattice models, provide a possible thermodynamic explanation for the TCM. 

While Eq.~\ref{eq:bridgman} is surprisingly accurate for water (and some other liquids) at near-standard conditions, it breaks down at high temperatures~\cite{Huber2012_WaterTC, Wagner2002_IAPWS1995}. 
Furthermore, advancing our results below, the magnitude of the Bridgman equation does not hold for molecular force fields of water even at near-standard conditions: it underestimates the TC by {$\sim$10-40\%} ($\sim$180\% for mW) at 300~K.
A popular approach is to replace the factor of 2.8 with a fitted coefficient, resulting in accurate predictions for a wide array of fluids, especially for monatomic and diatomic liquids~\cite{Khrapak2023_BridgmanTest}. The fitted coefficient is $\sim 1$ for monatomic fluids and generally increases with molecular complexity~\cite{Khrapak2023_BridgmanTest}; for water it is $\sim 3$, and $\sim 1.8$ at extreme conditions (1000-2000~K and 1.0-1.9~{g cm$^{-3}$}; up to 22~GPa)~\cite{Khrapak2023_BridgmanTest,Zhang2023_waterTCextreme}. This demonstrates that the TC is highly correlated with the speed of sound, and supports the use of the Bridgman equation to predict the position and existence of the TCM, which do not change when scaling Eq.~\ref{eq:bridgman} by a constant. 
Even so, while Eq.~\ref{eq:bridgman} reproduces the TCM, it is shifted by {$T_{\mathrm{max}(\lambda_\mathrm{B})} - T_{\mathrm{max}(\lambda)} \sim -70$~K} at near-atmospheric pressure, and $|T_{\mathrm{max}(\lambda_\mathrm{B})} - T_{\mathrm{max}(\lambda)}|$ increases with pressure~\cite{Huber2012_WaterTC}.

The question remains: to what extent is the TCM connected with the $\beta_T$ minimum? This hypothesis has eluded investigation by simulation studies because it is difficult to build an accurate model of water that does not reproduce the compressibility minimum, which reflects crucial aspects of the orientational correlations and tetrahedral order in liquid water. 
Furthermore, owing to the microscopic formulation of the heat flux~\cite{Boone2019_HeatFlux,Surblys2019_HeatFlux,Surblys2021_HeatFlux} and the presence of coupling effects~\cite{bresmeprl2008,Gittus2020}, it is still challenging to calculate TC from simulations, and the TCM has seldom been reported.
Advancing our discussion below, we identify six water models that reproduce the TCM, and crucially two that do not.

We perform extensive equilibrium and non-equilibrium molecular dynamics simulations for a diversity of water models: the rigid non-polarizable force fields TIP3P~\cite{Jorgensen1983_TIP3PandTIP4P}, SPC~\cite{Berendsen1981_SPC}, SPC/E~\cite{Berendsen1987_SPCE}, TIP4P/2005~\cite{Abascal2005_TIP4P2005} and TIP5P~\cite{Mahoney2000_TIP5P}; the flexible, polarizable and reactive force fields (ReaxFF),  water-2017~\cite{Zhang2017_2ndgen} and {CHON-2017\_weak}~\cite{Zhang2018_CHONweak}; and the highly coarse-grained monotonic water ``mW'' model~\cite{Molinero2009_mW} together with related Stillinger-Weber (SW) potentials~\cite{Stillinger1985_SWpotential}. 
For the rigid non-polarizable force fields, a cutoff of 13~{\AA} was used for the oxygen-oxygen Lennard-Jones (LJ) potential, and tail corrections~\cite{allen-tildesley-87} were \textit{not} applied in order for the isotropic equilibrium simulations to be fully consistent with the non-isotropic NEMD simulations.
The ReaxFF models use an interaction cutoff of {10~\AA}, which is part of the force field parameterization. 
All simulations were carried out using LAMMPS~\cite{LAMMPS}.
Thermodynamic response functions were calculated with simulations in the \textit{NPT} ensemble using the fluctuation relations and the equation of state. 
The TCs were calculated from boundary-driven non-equilibrium molecular dynamics (NEMD) simulations using Fourier's Law, $\bm{J}_q = - \lambda \nabla T$, where $\bm{J}_q$ is the heat flux and $\nabla T$ is the local temperature gradient. As demonstrated in our previous work, the temperature gradients {$\nabla T < 12$~{K nm$^{-1}$}} used in this work are well within the linear regime.~\cite{Gittus2021_ReaxFFWater} 
We note that the computation of TC via NEMD includes all possible coupling effects; in the case polar fluids such as water this includes the coupling between heat and polarization fluxes, which decreases the TC~\cite{bresmeprl2008,Gittus2020,Chapman2022_TPwater}.

\begin{figure*}[]
    \centering
 \begin{overpic}[width=0.8\textwidth,grid=False]{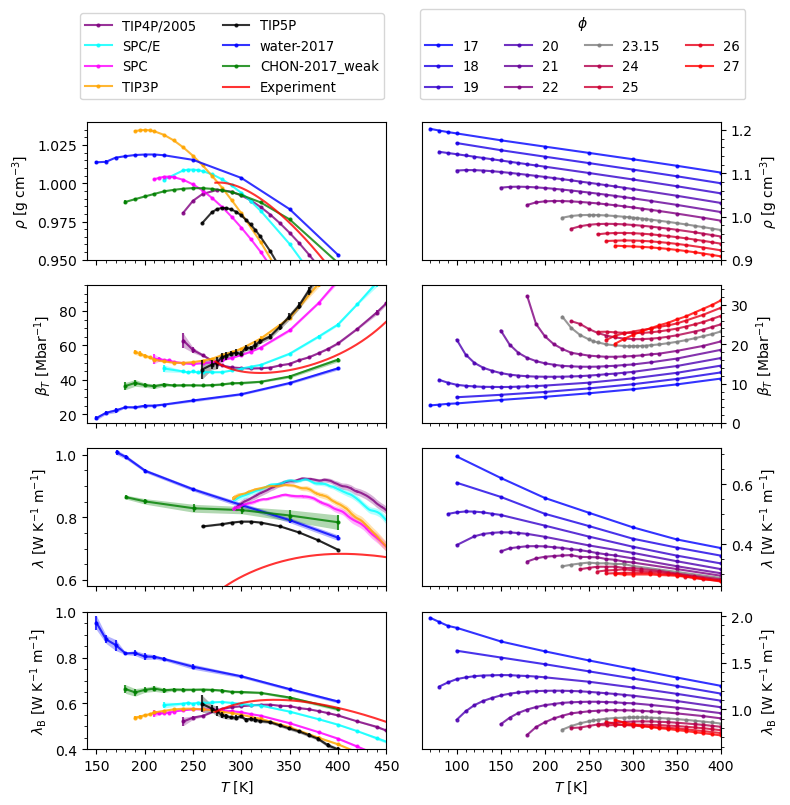}
        \put(0,86) {\textbf{\textsf{(a)}}}
        \put(95,86) {\textbf{\textsf{(b)}}}
        \put(45,82) {\textsf{(i)}}
        \put(12,62) {\textsf{(ii)}}
        \put(12,29) {\textsf{(iii)}}
        \put(43,21) {\textsf{(iv)}}

        \put(86.5,82) {\textsf{(i)}}
        \put(53.5,62) {\textsf{(ii)}}
        \put(53.5,29) {\textsf{(iii)}}
        \put(85,21) {\textsf{(iv)}}
    \end{overpic}
     \caption{Thermophysical properties of (a) atomistic force fields of water and (b) SW potentials as a function of temperature $T$ at constant pressure: the (i) density $\rho$, (ii) isothermal compressibility $\beta_T$, (iii) thermal conductivity $\lambda$, and (iv) Bridgman thermal conductivity $\lambda_\mathrm{B}$. Data corresponds to $\sim$10~bar (see Table~\ref{table:waterTC:Tex}).
     }
    \label{fig:waterTC:TCisobar}
\end{figure*}
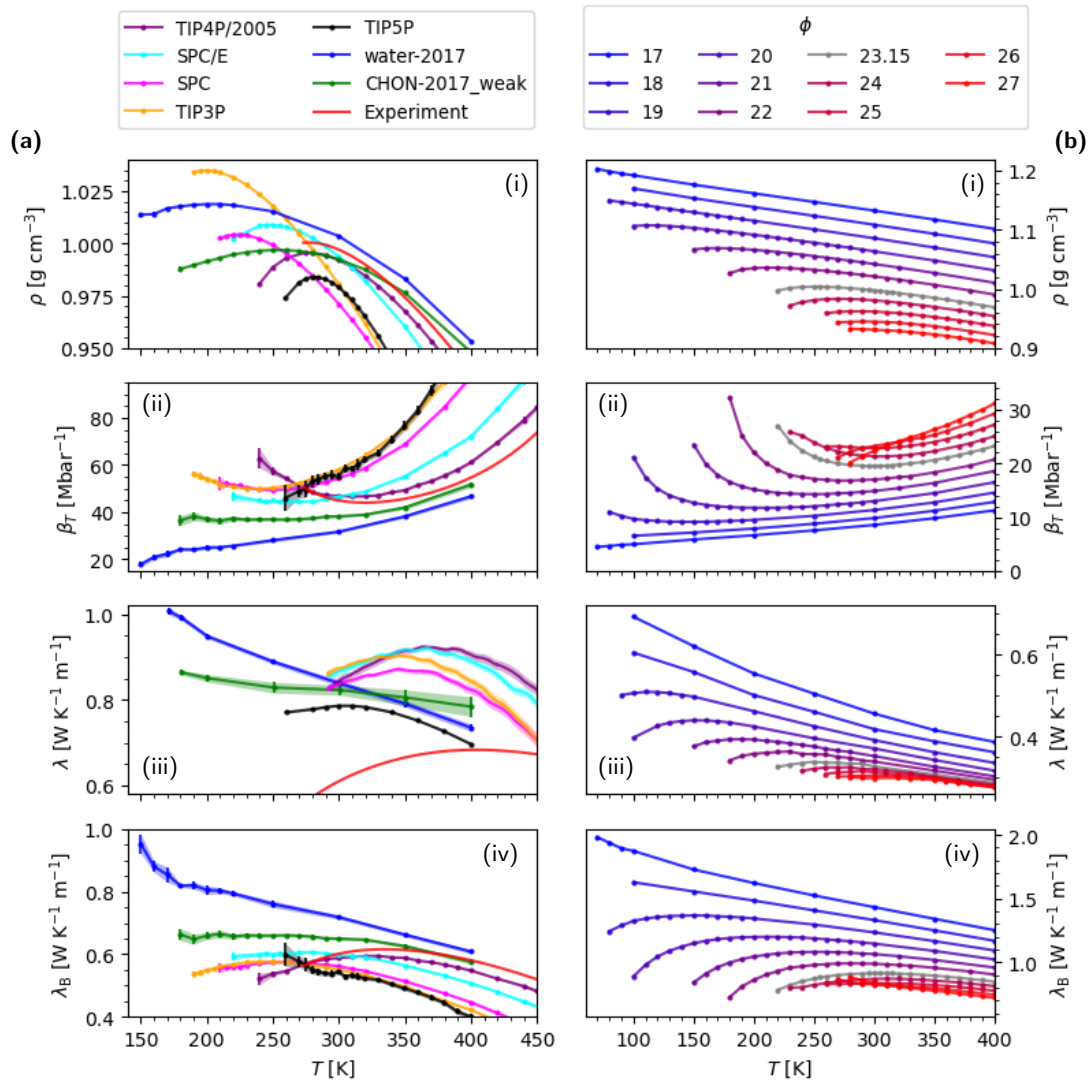

We show in {Fig.~\ref{fig:waterTC:TCisobar}(a)} the TC of the water models at constant pressure, alongside the experimental values. 
We target the 10~bar isobar since at 1~bar the TCM occurs above the boiling point and experimental data for super-heated water is not available. In our NEMD simulations the pressure of the stationary state is given by the average pressure tensor component in the direction of the heat flux. For the reactive force fields, TIP5P and the mW/SW potentials, a single $\lambda(P,T)$ point is calculated from each simulation, and the TC at 10~bar is calculated by interpolating $\lambda(P)$ at a given $T$. For TIP4P/2005, SPC/E, SPC and TIP3P, we use large NEMD simulations to extract $\lambda(T)$ as a continuous function of temperature, and therefore more accurately determine the temperature $T_{\mathrm{max}(\lambda)}$ of the TCM. These simulations correspond to pressures within $\pm 12$~bar of 10~bar as detailed in {Table~\ref{table:waterTC:Tex}}; deviations of only $\lesssim 10^{-3}$~{W K$^{-1}$ m$^{-1}$} from $\lambda(P/{\mathrm{bar}} = 10)$ are expected due to the small difference in pressure, which is smaller than the associated uncertainty. 
Consistent with the growing body of work~\cite{Ohara1999, Bedrov2000, Bresme2001, Zhang2005, Terao2005, William2007,Rosenbaum2007, Jiang2008, Kuang2010, Muscatello2011_JCP, Muscatello2011_ThermalPolarizationPCCP, Romer2012_waterTC, Mao2012, Sirk2013, Lee2014, Lee2019,Gittus2021_ReaxFFWater} demonstrating that empirical atomistic force fields typically overestimate the TC by {$\sim$10-50\%} at temperatures/pressures near 300~K and 1~bar, the atomistic force fields reported here systematically overestimate $\lambda$ by {$\sim$30-40\%} at near-standard conditions.

\begin{table*}[]
\small
\centering
\caption{The temperature $T_\mathrm{ex}$ at which extrema in thermophysical properties occur as a function of temperature at constant pressure $P$, as predicted by selected force fields. The temperatures of the density maximum $T_{\mathrm{max}(\rho)}$, isothermal compressibility minimum $T_{\mathrm{min}(\beta_T)}$, isentropic compressibility minimum  $T_{\mathrm{min}(\beta_S)}$, isentropic speed of sound maximum $T_{\mathrm{max}(c_S)}$, Bridgman thermal conductivity maximum $T_{\mathrm{max}(\lambda_\mathrm{B})}$, and thermal conductivity maximum $T_{\mathrm{max}(\lambda)}$. The thermal conductivity $\lambda$ and Bridgman thermal conductivity $\lambda_\mathrm{B}$ at 300~K are also reported.}
\label{table:waterTC:Tex}
\begin{threeparttable}
\begin{tabular}{c c c c c c c c c c}
\toprule
\hline 

Force Field & $P$ & $\lambda$ (300~K) & $\lambda_\mathrm{B}$ (300~K) & $T_{\mathrm{max}(\rho)}$ & $T_{\mathrm{min}(\beta_T)}$ & $T_{\mathrm{min}(\beta_S)}$ & $T_{\mathrm{max}(c_S)}$ &  $T_{\mathrm{max}(\lambda_\mathrm{B})}$ & $T_{\mathrm{max}(\lambda)}$ \\
 & [bar] & [W K$^{-1}$ m$^{-1}$] & [W K$^{-1}$ m$^{-1}$] & [K] &  [K] &  [K] &  [K] & [K] & [K] \\

\hline
TIP4P/2005 & 10 & 0.84(1) & 0.588(2) & 277(3) & 309(11) & 327(12) & 332(15) & 325(11) & 369(13) \\
SPC/E      & 12 & 0.87(1) & 0.598(2) & 248(4) & 261(17) & 272(15) & 280(13) & 269(15) & 361(14) \\
SPC        & 0 & 0.836(8) & 0.562(2) & 224(6) & 250(10) & 260(11) & 270(12) & 258(13) & 352(15) \\
TIP3P      & 22 & 0.876(5) & 0.551(2) & 199(4) & 238(9) & 247(10) & 254(12) & 245(11) & 343(18) \\
TIP5P      & 10 & 0.786(2) & 0.544(7) & 282(5) & None & None & None & None & 305(8) \\
water-2017 & 10 & 0.839(8) & 0.718(7) & 204(8) & None & None & None & None & None \\
{CHON-2017\_weak} & 10 & 0.82(1) & 0.650(6) & 255(6) & None & None & None & None & None \\
mW & 10 & 0.328(2) & 0.9134(6) & 251(1) & 299(5) & 305(5) & 309(6) & 304(6) & 253(8) \\

Exp.$^a$      & 10 & 0.610 & 0.602 & 277 & 320 & 337 & 347 & 334 & 404 \\

\hline
\end{tabular}
\begin{tablenotes}
\item[] $^a$ Experimental data are from, or calculated from, Refs.~\citenum{Huber2012_WaterTC}~\&~\citenum{Wagner2002_IAPWS1995}.
\end{tablenotes}
\end{threeparttable}
\end{table*}


The TCM can been inferred for TIP4P/2005, SPC/E and MCFM in previous NEMD simulation studies~\cite{Romer2012_waterTC,Muscatello2011_ThermalPolarizationPCCP}, but quantitative estimates of $T_{\mathrm{max}(\lambda)}$ were not reported. 
Studies using the Green-Kubo (GK) method have explicitly shown TCM at near-atmospheric pressures (exact pressure conditions were not specified) for TIP4P/2005 ($T_{\mathrm{max}(\lambda)} \sim 400$~K), SPC/E ($T_{\mathrm{max}(\lambda)}\sim 400$~K) and TIP4P ($T_{\mathrm{max}(\lambda)} \sim 350$~K).~\cite{Lee2014, Lee2019}
However, these studies report $\lambda$ values as low as {0.2~{W K$^{-1}$ m$^{-1}$}} at {240-250~K} in the supercooled liquid state, which is inconsistent with the {0.8-0.9~{W K$^{-1}$ m$^{-1}$}} from other~\cite{Bresme2014,English2014} GK predictions using the same models. 
Thus, to the best of our knowledge, we report the first reliable quantitative predictions of $T_{\mathrm{max}(\lambda)}$ using empirical force fields. 
Recently, a machine-learned potential trained at the quantum-mechanical DFT level with the SCAN functional predicted the TCM only when corrected for nuclear quantum effects (NQEs)~\cite{Xu2023_waterMLP_TC}.

SPC, SPC/E, TIP3P and TIP4P/2005 reproduce the TCM and the correct order of extrema, {$T_{\mathrm{max}(\lambda)} > T_{\mathrm{max}(c_S)} >  T_{\mathrm{min}(\beta_S)} > T_{\mathrm{max}(\lambda_\mathrm{B})} > T_{\mathrm{min}(\beta_T)} > T_{\mathrm{max}(\rho)} $}. 
The two models, water-2017 and {CHON-2017\_weak}, that fail to reproduce the TCM also fail to reproduce the extrema in the thermodynamic properties $\beta_T$, $\beta_S$, $c_S$ and $\lambda_\mathrm{B}$. 
These simulation results support the hypothesis that the TCM arises from the compressibility minimum. However, TIP5P does reproduce the TCM but not the four thermodynamic extrema, indicating other factors are at play in giving rise to the TCM. 
We will show below that investigating the behaviour of the TC  in tetrahedral liquids using the mW/SW potentials gives insight  into the behaviour of TIP5P.

The mW model underestimates the TC of water by {46\%} at 300~K, and is {$\sim$60\%} lower than the atomistic force fields. The TCM is also shifted to a lower temperature in the order of extrema: {$T_{\mathrm{max}(c_S)} >  T_{\mathrm{min}(\beta_S)} > T_{\mathrm{max}(\lambda_\mathrm{B})} > T_{\mathrm{min}(\beta_T)} > T_{\mathrm{max}(\lambda)} >  T_{\mathrm{max}(\rho)} $}. Decompositions of the microscopic heat flux in atomistic simulations show that Coulomb interactions~\cite{Matsubara2021_MolecHeatTransfer}, rotational intermolecular energy transfer~\cite{Ohara1999}, and the heat flux carried by the hydrogen atoms~\cite{Muscatello2011_ThermalPolarizationPCCP} are major contributors to water's TC. The mW model lacks all these mechanisms of heat transfer, explaining its lower TC.

\begin{figure}[]
    \centering
 \begin{overpic}[width=0.45\textwidth,grid=False]{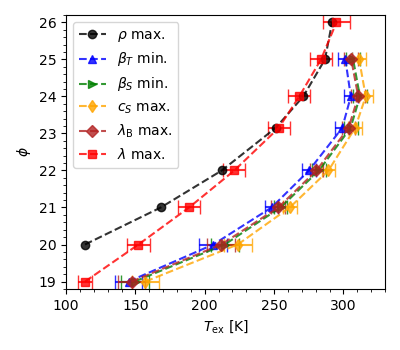}
    \end{overpic}
     \caption{The temperature $T_\mathrm{ex}$ at which extrema in thermophysical properties occur for the SW potentials at 10~bar. 
     }
    \label{fig:waterTC:TexSW}
\end{figure}

The mW model is a paramaterization of the SW potential for water. The SW potential has the form $\mathcal{V}_\mathrm{SW} = \mathcal{V}_2 + \phi \mathcal{V}_3$, where $\mathcal{V}_2$ and $\mathcal{V}_3$ are the potential energy contributions of the two- and three-body terms, respectively. $\mathcal{V}_3$ imposes an energetic penalty for deviations from the tetrahedral angle $\cos{\theta} = -1/3$ ($\theta \approx 109.5^{\circ}$) between triplets of particles. Increasing $\phi$ therefore increases the degree of tetrahedral order of the liquid.
By varying $\phi$, we systematically investigate the effect of tetrahedrality on the TCM.  
We show in {Fig.~\ref{fig:waterTC:TCisobar}(b)} $\rho$, $\beta_T$, $\lambda$ and $\lambda_\mathrm{B}$ for $17 \leq \phi \leq 27$ at 10~bar. 
At constant pressure (10~bar), the TC increases with $\phi$ due to the corresponding increase in density.
The behaviour of $\lambda_\mathrm{B}(T;\phi)$ mirrors that of $\lambda (T;\phi)$, justifying the interpretation with Eq.~\ref{eq:bridgman}.
We show in {Fig.~\ref{fig:waterTC:TexSW}} the temperature $T_\mathrm{ex}$ of the TCM and other extrema for the SW model.
$T_\mathrm{ex}$ features a maximum at $\phi \sim 24$ for $\beta_T$, $\beta_S$, $c_S$ and $\lambda_\mathrm{B}$, while $T_{\mathrm{max}(\rho)}$ and $T_{\mathrm{max}(\lambda)}$ increase monotonically with $\phi$.

Decreasing $\phi$ from the mW model ($\phi = 23.15$) to $\phi = 17$ interpolates between the behaviour of water and a simple liquid. 
First the density maximum is lost at $19 \leq \phi < 20$, followed by the TCM together with the extrema in  $\beta_T$, $\beta_S$, $c_S$ and $\lambda_\mathrm{B}$ at $18 \leq \phi < 19$, reaffirming the connection between the TCM and compressibility minimum. Thus, the SW model transitions from tetrahedral liquid behaviour at $\phi \sim 20$ to that of a simple fluid at $\phi \sim 18$.

Analogously, increasing $\phi$ from the mW model to $\phi = 27$ interpolates between the behaviour of water and that of highly tetrahedral materials such as carbon ($\phi = 26.2$~\cite{Barnard2002_SWcarbon}). In this case, the extrema in $\beta_T$, $\beta_S$ and $c_S$ disappear at $25 < \phi \leq 26$, before the TCM and density maximum, which are lost at $26 < \phi \leq 27$. 
For $25 \lesssim \phi \lesssim 27$ where there is no compressibility minimum, the qualitative interpretation of the Bridgman equation suggests the TCM arises from the density maximum and the monotonically increasing $\beta_S (T)$ shifts $T_{\mathrm{max}(\lambda)}$ to a lower temperature; consistent with this we observe $T_{\mathrm{max}(\lambda)} \approx T_{\mathrm{max}(\rho)}$ for $\phi = 26$. However, $\phi = 26$ does not feature a $\lambda_\mathrm{B}$ maximum, reflecting that while Eq.~\ref{eq:bridgman} is instructive, it does not reproduce the exact balance between $\rho(T)$ and $\beta_T (T)$ in determining the TC.  

We note that at $\phi=17,18,27$ the thermal expansion coefficient $\alpha_P$ features a minimum with $\mathrm{min}(\alpha_P) > 0$.  
For $\phi=19$ extrapolating $\alpha_P$ to $T = 0$ gives $\alpha_P > 0$ for the entire temperature range. This implies that the density maximum is not observed at lower temperatures at these values of $\phi$.

From the above analysis we identify five regimes for the TCM of tetrahedral liquids: (1) the simple liquid regime at $\phi \lesssim 18$ where all six extrema do not exist; (2) the compressibility regime at $18 \lesssim \phi \lesssim 20$ where the TCM arises solely from the $\beta_T$ (or equivalently $\beta_S$) minimum; (3) the mixed regime, $20 \lesssim \phi \lesssim 25$ where both the density maximum and compressibility minimum can contribute to the TCM; (4) the density regime at $25 \lesssim \phi \lesssim 27$ where the TCM arises solely from the density maximum; and (5) the highly tetrahedral regime at $\phi \gtrsim 27$ where all six extrema do not exist.

\begin{figure}[]
    \centering
 \begin{overpic}[width=0.45\textwidth,grid=False]{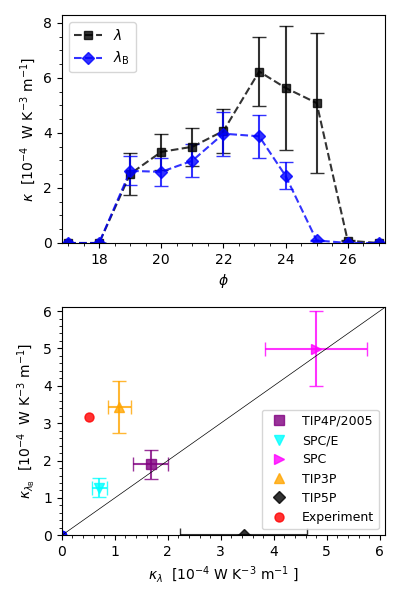}
        \put(0,97) {\textsf{(a)}}
        \put(0,49) {\textsf{(b)}}
    \end{overpic}
     \caption{(a) Curvature ${\kappa}$ of the thermal conductivity $\lambda$ and Bridgman thermal conductivity $\lambda_\mathrm{B}$ maxima as a function of $\phi$ for the SW potential. (b) $\kappa_\lambda$ \textit{vs.} $\kappa_{\lambda_\mathrm{B}}$ for selected atomistic force fields; $\kappa_x$ is the curvature of $x$. The solid line shows $\kappa_{\lambda_\mathrm{B}} = \kappa_\lambda$.
     }
    \label{fig:waterTC:curv}
\end{figure}

The curvature $\kappa$ of the TCM provides a measure of its \enquote{intensity}, with $\kappa = 0$ corresponding to the loss of the maximum.
In {Fig~\ref{fig:waterTC:curv}(a)} we show the curvature of the $\lambda$ and $\lambda_\mathrm{B}$ maxima for the SW model as a function of $\phi$. $\kappa$ was determined by fitting a cubic function to the data close to the maximum.
$\kappa_\lambda$ increases from 0 in the simple liquid regime, reaches a maximum in the mixed regime at $\phi \sim 23$ corresponding to water, then decreases before vanishing at highly tetrahedral values of $\phi$. The Bridgman equation reproduces this trend in curvature, further supporting its ability to qualitatively describe the TCM.

The mW model exists in the mixed regime due to the shift in $T_{\mathrm{max}(\lambda)}$ of {$\sim -$150~K} compared to experiment that places its TCM close to its density maximum, $T_{\mathrm{max}(\lambda)} \sim T_{\mathrm{max}(\rho)}$. This shift may be attributed to the different heat transfer mechanism that is also responsible for the mW model's lower TC.
However, {$T_{\mathrm{max}(\lambda)} > T_{\mathrm{min}(\beta_S)} \gtrsim T_{\mathrm{max}(\lambda_\mathrm{B})} \gtrsim T_{\mathrm{min}(\beta_T)} > T_{\mathrm{max}(\rho)} $} for water and all the atomistic force fields that reproduce all the extrema, such that the $\lambda_\mathrm{B}$ maximum and TCM originate solely from the compressibility minimum. To further support the correlation between the $\lambda_\mathrm{B}$ and $\lambda$ maxima in this case, we show in {Fig.~\ref{fig:waterTC:curv}(a)} that the curvatures, $\kappa_\lambda$ and $\kappa_{\lambda_\mathrm{B}}$, of these maxima are correlated. 
The exception to this is TIP5P, which features a maximum in $\lambda$ but not $\lambda_\mathrm{B}$; this is the same scenario as $\phi = 26$ for the SW model, for which the TCM is in the density regime. One may similarly interpret the TIP5P TCM as arising from the density maximum; indeed, with {$T_{\mathrm{max}(\lambda)} - T_{\mathrm{max}(\rho)} = (23 \pm 9)$~K}, the TCM is closer to the density maximum by $\sim${70-120}~K than for the other atomistic force field, and by $\sim 100$~K compared to experiment.

In conclusion, we present the the most accurate calculations of water's TCM as predicted by selected molecular force fields. By examining why models succeed or fail to reproduce the TCM, and interpreting our results using theoretical quasi-lattice models, namely the Bridgman equation, we show that water's TCM is connected to its compressibility minimum. 
The $\beta_T$ minimum has been studied extensively, and microscopic explanations have been proposed by now widely accepted two-state models in which two structural motifs coexist in water: the low-density (enthalpy favoured through tetrahedral H-bonding) and and high-density (entropy favoured) states~\cite{Skinner2014,Huang2011,Schlesinger2016,Nilsson2015,Tanaka2014,Vedamuthu1995}. 
An increased concentration of tetrahedral structures (increasing $\beta_T$) upon cooling competes with effect of increasing density (decreasing $\beta_T$), leading to a minimum. Thus, we additionally provide a microscopic explanation for the TCM by connecting it with the $\beta_T$ minimum.

Using SW potentials for tetrahedral liquids, we investigate the TCM as a function of tetrahedrality, interpolating between the behaviour of simple liquids and highly tetrahedral materials such as carbon.  
Together with the two vanishing limits at low/high tetrahedrality, we identify three different regimes for the TCM: when it originates from either the compressibility minimum or density maximum, or both. Our results indicate that the TCM in real water exists in a \enquote{Goldilocks Zone} of tetrahedral order. We show that the maximum \enquote{intensity} (i.e., curvature) of the TCM, occurs at the tetrahedrality of liquid water, suggesting that the TC of water has a maximally anomalous temperature dependence compared to other tetrahedral liquids.
We provide a thermodynamic explanation for the TCM of not only water, but tetrahedral liquids in general.

\begin{acknowledgments}
We thank the Leverhulme Trust for Grant No. RPG-2018-384. 
We gratefully acknowledge a PhD studentship (Project Reference 2135626) for O.R.G. sponsored by ICL's Chemistry Doctoral Scholarship Award, funded by the EPSRC Doctoral Training Partnership Account (EP/N509486/1). 
We acknowledge the ICL RCS High Performance Computing facility and the UK Materials and Molecular Modelling Hub for computational resources, partially funded by the EPSRC (Grant Nos. EP/P020194/1
and EP/T022213/1).
\end{acknowledgments}

\bibliography{refs}

\begin{thebibliography}{55}%
\makeatletter
\providecommand \@ifxundefined [1]{%
 \@ifx{#1\undefined}
}%
\providecommand \@ifnum [1]{%
 \ifnum #1\expandafter \@firstoftwo
 \else \expandafter \@secondoftwo
 \fi
}%
\providecommand \@ifx [1]{%
 \ifx #1\expandafter \@firstoftwo
 \else \expandafter \@secondoftwo
 \fi
}%
\providecommand \natexlab [1]{#1}%
\providecommand \enquote  [1]{``#1''}%
\providecommand \bibnamefont  [1]{#1}%
\providecommand \bibfnamefont [1]{#1}%
\providecommand \citenamefont [1]{#1}%
\providecommand \href@noop [0]{\@secondoftwo}%
\providecommand \href [0]{\begingroup \@sanitize@url \@href}%
\providecommand \@href[1]{\@@startlink{#1}\@@href}%
\providecommand \@@href[1]{\endgroup#1\@@endlink}%
\providecommand \@sanitize@url [0]{\catcode `\\12\catcode `\$12\catcode `\&12\catcode `\#12\catcode `\^12\catcode `\_12\catcode `\%12\relax}%
\providecommand \@@startlink[1]{}%
\providecommand \@@endlink[0]{}%
\providecommand \url  [0]{\begingroup\@sanitize@url \@url }%
\providecommand \@url [1]{\endgroup\@href {#1}{\urlprefix }}%
\providecommand \urlprefix  [0]{URL }%
\providecommand \Eprint [0]{\href }%
\providecommand \doibase [0]{https://doi.org/}%
\providecommand \selectlanguage [0]{\@gobble}%
\providecommand \bibinfo  [0]{\@secondoftwo}%
\providecommand \bibfield  [0]{\@secondoftwo}%
\providecommand \translation [1]{[#1]}%
\providecommand \BibitemOpen [0]{}%
\providecommand \bibitemStop [0]{}%
\providecommand \bibitemNoStop [0]{.\EOS\space}%
\providecommand \EOS [0]{\spacefactor3000\relax}%
\providecommand \BibitemShut  [1]{\csname bibitem#1\endcsname}%
\let\auto@bib@innerbib\@empty
\bibitem [{\citenamefont {Huber}\ \emph {et~al.}(2012)\citenamefont {Huber}, \citenamefont {Perkins}, \citenamefont {Friend}, \citenamefont {Sengers}, \citenamefont {Assael}, \citenamefont {Metaxa}, \citenamefont {Miyagawa}, \citenamefont {Hellmann},\ and\ \citenamefont {Vogel}}]{Huber2012_WaterTC}%
  \BibitemOpen
  \bibfield  {author} {\bibinfo {author} {\bibfnamefont {M.~L.}\ \bibnamefont {Huber}}, \bibinfo {author} {\bibfnamefont {R.~A.}\ \bibnamefont {Perkins}}, \bibinfo {author} {\bibfnamefont {D.~G.}\ \bibnamefont {Friend}}, \bibinfo {author} {\bibfnamefont {J.~V.}\ \bibnamefont {Sengers}}, \bibinfo {author} {\bibfnamefont {M.~J.}\ \bibnamefont {Assael}}, \bibinfo {author} {\bibfnamefont {I.~N.}\ \bibnamefont {Metaxa}}, \bibinfo {author} {\bibfnamefont {K.}~\bibnamefont {Miyagawa}}, \bibinfo {author} {\bibfnamefont {R.}~\bibnamefont {Hellmann}},\ and\ \bibinfo {author} {\bibfnamefont {E.}~\bibnamefont {Vogel}},\ }\bibfield  {title} {\bibinfo {title} {New international formulation for the thermal conductivity of h2o},\ }\href {https://doi.org/10.1063/1.4738955} {\bibfield  {journal} {\bibinfo  {journal} {J. Phys. Chem. Ref. Data}\ }\textbf {\bibinfo {volume} {41}},\ \bibinfo {pages} {033102} (\bibinfo {year} {2012})}\BibitemShut {NoStop}%
\bibitem [{\citenamefont {Bridgman}(1923)}]{Bridgman1923_TCeq}%
  \BibitemOpen
  \bibfield  {author} {\bibinfo {author} {\bibfnamefont {P.~W.}\ \bibnamefont {Bridgman}},\ }\bibfield  {title} {\bibinfo {title} {The thermal conductivity of liquids},\ }\href {https://doi.org/10.1073/pnas.9.10.341} {\bibfield  {journal} {\bibinfo  {journal} {Proc. Natl. Acad. Sci. U.S.A.}\ }\textbf {\bibinfo {volume} {9}},\ \bibinfo {pages} {341} (\bibinfo {year} {1923})},\ \Eprint {https://arxiv.org/abs/https://www.pnas.org/doi/pdf/10.1073/pnas.9.10.341} {https://www.pnas.org/doi/pdf/10.1073/pnas.9.10.341} \BibitemShut {NoStop}%
\bibitem [{\citenamefont {McLaughlin}(1964)}]{McLaughlin1964_TC}%
  \BibitemOpen
  \bibfield  {author} {\bibinfo {author} {\bibfnamefont {E.}~\bibnamefont {McLaughlin}},\ }\bibfield  {title} {\bibinfo {title} {The thermal conductivity of liquids and dense gases},\ }\href {https://doi.org/10.1021/cr60230a003} {\bibfield  {journal} {\bibinfo  {journal} {Chem. Rev.}\ }\textbf {\bibinfo {volume} {64}},\ \bibinfo {pages} {389} (\bibinfo {year} {1964})},\ \Eprint {https://arxiv.org/abs/https://doi.org/10.1021/cr60230a003} {https://doi.org/10.1021/cr60230a003} \BibitemShut {NoStop}%
\bibitem [{\citenamefont {Chen}(2021)}]{Chen2020_ThermophysicsLiquids}%
  \BibitemOpen
  \bibfield  {author} {\bibinfo {author} {\bibfnamefont {G.}~\bibnamefont {Chen}},\ }\bibfield  {title} {\bibinfo {title} {{Perspectives on Molecular-Level Understanding of Thermophysics of Liquids and Future Research Directions}},\ }\bibfield  {journal} {\bibinfo  {journal} {J. Heat Transfer}\ }\textbf {\bibinfo {volume} {144}},\ \href {https://doi.org/10.1115/1.4052657} {10.1115/1.4052657} (\bibinfo {year} {2021}),\ \bibinfo {note} {010801},\ \Eprint {https://arxiv.org/abs/https://asmedigitalcollection.asme.org/heattransfer/article-pdf/144/1/010801/6804357/ht\_144\_01\_010801.pdf} {https://asmedigitalcollection.asme.org/heattransfer/article-pdf/144/1/010801/6804357/ht\_144\_01\_010801.pdf} \BibitemShut {NoStop}%
\bibitem [{\citenamefont {Xi}\ \emph {et~al.}(2020)\citenamefont {Xi}, \citenamefont {Zhong}, \citenamefont {He}, \citenamefont {Xu}, \citenamefont {Nakayama}, \citenamefont {Wang}, \citenamefont {Liu}, \citenamefont {Zhou},\ and\ \citenamefont {Li}}]{Xi2020_TCmodel}%
  \BibitemOpen
  \bibfield  {author} {\bibinfo {author} {\bibfnamefont {Q.}~\bibnamefont {Xi}}, \bibinfo {author} {\bibfnamefont {J.}~\bibnamefont {Zhong}}, \bibinfo {author} {\bibfnamefont {J.}~\bibnamefont {He}}, \bibinfo {author} {\bibfnamefont {X.}~\bibnamefont {Xu}}, \bibinfo {author} {\bibfnamefont {T.}~\bibnamefont {Nakayama}}, \bibinfo {author} {\bibfnamefont {Y.}~\bibnamefont {Wang}}, \bibinfo {author} {\bibfnamefont {J.}~\bibnamefont {Liu}}, \bibinfo {author} {\bibfnamefont {J.}~\bibnamefont {Zhou}},\ and\ \bibinfo {author} {\bibfnamefont {B.}~\bibnamefont {Li}},\ }\bibfield  {title} {\bibinfo {title} {A ubiquitous thermal conductivity formula for liquids, polymer glass, and amorphous solids},\ }\href {https://doi.org/10.1088/0256-307x/37/10/104401} {\bibfield  {journal} {\bibinfo  {journal} {Chinese Phys. Lett.}\ }\textbf {\bibinfo {volume} {37}},\ \bibinfo {pages} {104401} (\bibinfo {year} {2020})}\BibitemShut {NoStop}%
\bibitem [{\citenamefont {Zhao}\ \emph {et~al.}(2021)\citenamefont {Zhao}, \citenamefont {Wingert}, \citenamefont {Chen},\ and\ \citenamefont {Garay}}]{Zhao2021_liquidTC}%
  \BibitemOpen
  \bibfield  {author} {\bibinfo {author} {\bibfnamefont {A.~Z.}\ \bibnamefont {Zhao}}, \bibinfo {author} {\bibfnamefont {M.~C.}\ \bibnamefont {Wingert}}, \bibinfo {author} {\bibfnamefont {R.}~\bibnamefont {Chen}},\ and\ \bibinfo {author} {\bibfnamefont {J.~E.}\ \bibnamefont {Garay}},\ }\bibfield  {title} {\bibinfo {title} {{Phonon gas model for thermal conductivity of dense, strongly interacting liquids}},\ }\href {https://doi.org/10.1063/5.0040734} {\bibfield  {journal} {\bibinfo  {journal} {Journal of Applied Physics}\ }\textbf {\bibinfo {volume} {129}},\ \bibinfo {pages} {235101} (\bibinfo {year} {2021})},\ \Eprint {https://arxiv.org/abs/https://pubs.aip.org/aip/jap/article-pdf/doi/10.1063/5.0040734/15264628/235101\_1\_online.pdf} {https://pubs.aip.org/aip/jap/article-pdf/doi/10.1063/5.0040734/15264628/235101\_1\_online.pdf} \BibitemShut {NoStop}%
\bibitem [{\citenamefont {Khrapak}(2021)}]{Khrapak2021_liquidTC}%
  \BibitemOpen
  \bibfield  {author} {\bibinfo {author} {\bibfnamefont {S.~A.}\ \bibnamefont {Khrapak}},\ }\bibfield  {title} {\bibinfo {title} {Vibrational model of thermal conduction for fluids with soft interactions},\ }\href {https://doi.org/10.1103/PhysRevE.103.013207} {\bibfield  {journal} {\bibinfo  {journal} {Phys. Rev. E}\ }\textbf {\bibinfo {volume} {103}},\ \bibinfo {pages} {013207} (\bibinfo {year} {2021})}\BibitemShut {NoStop}%
\bibitem [{\citenamefont {Khrapak}(2023)}]{Khrapak2023_BridgmanTest}%
  \BibitemOpen
  \bibfield  {author} {\bibinfo {author} {\bibfnamefont {S.~A.}\ \bibnamefont {Khrapak}},\ }\bibfield  {title} {\bibinfo {title} {Bridgman formula for the thermal conductivity of atomic and molecular liquids},\ }\href {https://doi.org/https://doi.org/10.1016/j.molliq.2023.121786} {\bibfield  {journal} {\bibinfo  {journal} {J. Mol. Liq.}\ }\textbf {\bibinfo {volume} {381}},\ \bibinfo {pages} {121786} (\bibinfo {year} {2023})}\BibitemShut {NoStop}%
\bibitem [{\citenamefont {Zhang}\ \emph {et~al.}(2023)\citenamefont {Zhang}, \citenamefont {Puligheddu}, \citenamefont {Zhang}, \citenamefont {Car},\ and\ \citenamefont {Galli}}]{Zhang2023_waterTCextreme}%
  \BibitemOpen
  \bibfield  {author} {\bibinfo {author} {\bibfnamefont {C.}~\bibnamefont {Zhang}}, \bibinfo {author} {\bibfnamefont {M.}~\bibnamefont {Puligheddu}}, \bibinfo {author} {\bibfnamefont {L.}~\bibnamefont {Zhang}}, \bibinfo {author} {\bibfnamefont {R.}~\bibnamefont {Car}},\ and\ \bibinfo {author} {\bibfnamefont {G.}~\bibnamefont {Galli}},\ }\bibfield  {title} {\bibinfo {title} {Thermal conductivity of water at extreme conditions},\ }\href {https://doi.org/10.1021/acs.jpcb.3c02972} {\bibfield  {journal} {\bibinfo  {journal} {J. Phys. Chem. B}\ }\textbf {\bibinfo {volume} {127}},\ \bibinfo {pages} {7011} (\bibinfo {year} {2023})},\ \bibinfo {note} {pMID: 37524047},\ \Eprint {https://arxiv.org/abs/https://doi.org/10.1021/acs.jpcb.3c02972} {https://doi.org/10.1021/acs.jpcb.3c02972} \BibitemShut {NoStop}%
\bibitem [{\citenamefont {Wagner}\ and\ \citenamefont {Pruß}(2002)}]{Wagner2002_IAPWS1995}%
  \BibitemOpen
  \bibfield  {author} {\bibinfo {author} {\bibfnamefont {W.}~\bibnamefont {Wagner}}\ and\ \bibinfo {author} {\bibfnamefont {A.}~\bibnamefont {Pruß}},\ }\bibfield  {title} {\bibinfo {title} {The iapws formulation 1995 for the thermodynamic properties of ordinary water substance for general and scientific use},\ }\href {https://doi.org/10.1063/1.1461829} {\bibfield  {journal} {\bibinfo  {journal} {J. Phys. Chem. Ref. Data}\ }\textbf {\bibinfo {volume} {31}},\ \bibinfo {pages} {387} (\bibinfo {year} {2002})}\BibitemShut {NoStop}%
\bibitem [{\citenamefont {Boone}\ \emph {et~al.}(2019)\citenamefont {Boone}, \citenamefont {Babaei},\ and\ \citenamefont {Wilmer}}]{Boone2019_HeatFlux}%
  \BibitemOpen
  \bibfield  {author} {\bibinfo {author} {\bibfnamefont {P.}~\bibnamefont {Boone}}, \bibinfo {author} {\bibfnamefont {H.}~\bibnamefont {Babaei}},\ and\ \bibinfo {author} {\bibfnamefont {C.~E.}\ \bibnamefont {Wilmer}},\ }\bibfield  {title} {\bibinfo {title} {Heat flux for many-body interactions: Corrections to lammps},\ }\href {https://doi.org/doi: 10.1021/acs.jctc.9b00252} {\bibfield  {journal} {\bibinfo  {journal} {J. Chem. Theory Comput.}\ }\textbf {\bibinfo {volume} {15}},\ \bibinfo {pages} {5579} (\bibinfo {year} {2019})}\BibitemShut {NoStop}%
\bibitem [{\citenamefont {Surblys}\ \emph {et~al.}(2019)\citenamefont {Surblys}, \citenamefont {Matsubara}, \citenamefont {Kikugawa},\ and\ \citenamefont {Ohara}}]{Surblys2019_HeatFlux}%
  \BibitemOpen
  \bibfield  {author} {\bibinfo {author} {\bibfnamefont {D.}~\bibnamefont {Surblys}}, \bibinfo {author} {\bibfnamefont {H.}~\bibnamefont {Matsubara}}, \bibinfo {author} {\bibfnamefont {G.}~\bibnamefont {Kikugawa}},\ and\ \bibinfo {author} {\bibfnamefont {T.}~\bibnamefont {Ohara}},\ }\bibfield  {title} {\bibinfo {title} {Application of atomic stress to compute heat flux via molecular dynamics for systems with many-body interactions},\ }\href {https://doi.org/10.1103/PhysRevE.99.051301} {\bibfield  {journal} {\bibinfo  {journal} {Phys. Rev. E}\ }\textbf {\bibinfo {volume} {99}},\ \bibinfo {pages} {051301} (\bibinfo {year} {2019})}\BibitemShut {NoStop}%
\bibitem [{\citenamefont {Surblys}\ \emph {et~al.}(2021)\citenamefont {Surblys}, \citenamefont {Matsubara}, \citenamefont {Kikugawa},\ and\ \citenamefont {Ohara}}]{Surblys2021_HeatFlux}%
  \BibitemOpen
  \bibfield  {author} {\bibinfo {author} {\bibfnamefont {D.}~\bibnamefont {Surblys}}, \bibinfo {author} {\bibfnamefont {H.}~\bibnamefont {Matsubara}}, \bibinfo {author} {\bibfnamefont {G.}~\bibnamefont {Kikugawa}},\ and\ \bibinfo {author} {\bibfnamefont {T.}~\bibnamefont {Ohara}},\ }\bibfield  {title} {\bibinfo {title} {Methodology and meaning of computing heat flux via atomic stress in systems with constraint dynamics},\ }\href {https://doi.org/10.1063/5.0070930} {\bibfield  {journal} {\bibinfo  {journal} {J. Appl. Phys.}\ }\textbf {\bibinfo {volume} {130}},\ \bibinfo {pages} {215104} (\bibinfo {year} {2021})},\ \Eprint {https://arxiv.org/abs/https://doi.org/10.1063/5.0070930} {https://doi.org/10.1063/5.0070930} \BibitemShut {NoStop}%
\bibitem [{\citenamefont {Bresme}\ \emph {et~al.}(2008)\citenamefont {Bresme}, \citenamefont {Lervik}, \citenamefont {Bedeaux},\ and\ \citenamefont {Kjelstrup}}]{bresmeprl2008}%
  \BibitemOpen
  \bibfield  {author} {\bibinfo {author} {\bibfnamefont {F.}~\bibnamefont {Bresme}}, \bibinfo {author} {\bibfnamefont {A.}~\bibnamefont {Lervik}}, \bibinfo {author} {\bibfnamefont {D.}~\bibnamefont {Bedeaux}},\ and\ \bibinfo {author} {\bibfnamefont {S.}~\bibnamefont {Kjelstrup}},\ }\bibfield  {title} {\bibinfo {title} {Water polarization under thermal gradients},\ }\href {https://doi.org/10.1103/PhysRevLett.101.020602} {\bibfield  {journal} {\bibinfo  {journal} {Phys. Rev. Lett.}\ }\textbf {\bibinfo {volume} {101}},\ \bibinfo {pages} {020602} (\bibinfo {year} {2008})}\BibitemShut {NoStop}%
\bibitem [{\citenamefont {Gittus}\ \emph {et~al.}(2020)\citenamefont {Gittus}, \citenamefont {Albella},\ and\ \citenamefont {Bresme}}]{Gittus2020}%
  \BibitemOpen
  \bibfield  {author} {\bibinfo {author} {\bibfnamefont {O.~R.}\ \bibnamefont {Gittus}}, \bibinfo {author} {\bibfnamefont {P.}~\bibnamefont {Albella}},\ and\ \bibinfo {author} {\bibfnamefont {F.}~\bibnamefont {Bresme}},\ }\bibfield  {title} {\bibinfo {title} {Polarization of acetonitrile under thermal fields via non-equilibrium molecular dynamics simulations},\ }\href {https://doi.org/10.1063/5.0025148} {\bibfield  {journal} {\bibinfo  {journal} {J. Chem. Phys.}\ }\textbf {\bibinfo {volume} {153}},\ \bibinfo {pages} {204503} (\bibinfo {year} {2020})}\BibitemShut {NoStop}%
\bibitem [{\citenamefont {Jorgensen}\ \emph {et~al.}(1983)\citenamefont {Jorgensen}, \citenamefont {Chandrasekhar}, \citenamefont {Madura}, \citenamefont {Impey},\ and\ \citenamefont {Klein}}]{Jorgensen1983_TIP3PandTIP4P}%
  \BibitemOpen
  \bibfield  {author} {\bibinfo {author} {\bibfnamefont {W.~L.}\ \bibnamefont {Jorgensen}}, \bibinfo {author} {\bibfnamefont {J.}~\bibnamefont {Chandrasekhar}}, \bibinfo {author} {\bibfnamefont {J.~D.}\ \bibnamefont {Madura}}, \bibinfo {author} {\bibfnamefont {R.~W.}\ \bibnamefont {Impey}},\ and\ \bibinfo {author} {\bibfnamefont {M.~L.}\ \bibnamefont {Klein}},\ }\bibfield  {title} {\bibinfo {title} {Comparison of simple potential functions for simulating liquid water},\ }\href {https://doi.org/10.1063/1.445869} {\bibfield  {journal} {\bibinfo  {journal} {J. Chem. Phys.}\ }\textbf {\bibinfo {volume} {79}},\ \bibinfo {pages} {926} (\bibinfo {year} {1983})},\ \Eprint {https://arxiv.org/abs/https://doi.org/10.1063/1.445869} {https://doi.org/10.1063/1.445869} \BibitemShut {NoStop}%
\bibitem [{\citenamefont {Berendsen}\ \emph {et~al.}(1981)\citenamefont {Berendsen}, \citenamefont {Postma}, \citenamefont {van Gunsteren},\ and\ \citenamefont {Hermans}}]{Berendsen1981_SPC}%
  \BibitemOpen
  \bibfield  {author} {\bibinfo {author} {\bibfnamefont {H.~J.~C.}\ \bibnamefont {Berendsen}}, \bibinfo {author} {\bibfnamefont {J.~P.~M.}\ \bibnamefont {Postma}}, \bibinfo {author} {\bibfnamefont {W.~F.}\ \bibnamefont {van Gunsteren}},\ and\ \bibinfo {author} {\bibfnamefont {J.}~\bibnamefont {Hermans}},\ }\bibinfo {title} {Interaction models for water in relation to protein hydration},\ in\ \href {https://doi.org/10.1007/978-94-015-7658-1_21} {\emph {\bibinfo {booktitle} {Intermolecular Forces: Proceedings of the Fourteenth Jerusalem Symposium on Quantum Chemistry and Biochemistry Held in Jerusalem, Israel, April 13--16, 1981}}},\ \bibinfo {editor} {edited by\ \bibinfo {editor} {\bibfnamefont {B.}~\bibnamefont {Pullman}}}\ (\bibinfo  {publisher} {Springer Netherlands},\ \bibinfo {address} {Dordrecht},\ \bibinfo {year} {1981})\ pp.\ \bibinfo {pages} {331--342}\BibitemShut {NoStop}%
\bibitem [{\citenamefont {Berendsen}\ \emph {et~al.}(1987)\citenamefont {Berendsen}, \citenamefont {Grigera},\ and\ \citenamefont {Straatsma}}]{Berendsen1987_SPCE}%
  \BibitemOpen
  \bibfield  {author} {\bibinfo {author} {\bibfnamefont {H.~J.~C.}\ \bibnamefont {Berendsen}}, \bibinfo {author} {\bibfnamefont {J.~R.}\ \bibnamefont {Grigera}},\ and\ \bibinfo {author} {\bibfnamefont {T.~P.}\ \bibnamefont {Straatsma}},\ }\bibfield  {title} {\bibinfo {title} {The missing term in effective pair potentials},\ }\href {https://doi.org/10.1021/j100308a038} {\bibfield  {journal} {\bibinfo  {journal} {J. Phys. Chem.}\ }\textbf {\bibinfo {volume} {91}},\ \bibinfo {pages} {6269} (\bibinfo {year} {1987})}\BibitemShut {NoStop}%
\bibitem [{\citenamefont {Abascal}\ and\ \citenamefont {Vega}(2005)}]{Abascal2005_TIP4P2005}%
  \BibitemOpen
  \bibfield  {author} {\bibinfo {author} {\bibfnamefont {J.~L.~F.}\ \bibnamefont {Abascal}}\ and\ \bibinfo {author} {\bibfnamefont {C.}~\bibnamefont {Vega}},\ }\bibfield  {title} {\bibinfo {title} {A general purpose model for the condensed phases of water: Tip4p/2005},\ }\href {https://doi.org/10.1063/1.2121687} {\bibfield  {journal} {\bibinfo  {journal} {J. Chem. Phys.}\ }\textbf {\bibinfo {volume} {123}},\ \bibinfo {pages} {234505} (\bibinfo {year} {2005})}\BibitemShut {NoStop}%
\bibitem [{\citenamefont {Mahoney}\ and\ \citenamefont {Jorgensen}(2000)}]{Mahoney2000_TIP5P}%
  \BibitemOpen
  \bibfield  {author} {\bibinfo {author} {\bibfnamefont {M.~W.}\ \bibnamefont {Mahoney}}\ and\ \bibinfo {author} {\bibfnamefont {W.~L.}\ \bibnamefont {Jorgensen}},\ }\bibfield  {title} {\bibinfo {title} {{A five-site model for liquid water and the reproduction of the density anomaly by rigid, nonpolarizable potential functions}},\ }\href {https://doi.org/10.1063/1.481505} {\bibfield  {journal} {\bibinfo  {journal} {J. Chem. Phys.}\ }\textbf {\bibinfo {volume} {112}},\ \bibinfo {pages} {8910} (\bibinfo {year} {2000})},\ \Eprint {https://arxiv.org/abs/https://pubs.aip.org/aip/jcp/article-pdf/112/20/8910/10806193/8910\_1\_online.pdf} {https://pubs.aip.org/aip/jcp/article-pdf/112/20/8910/10806193/8910\_1\_online.pdf} \BibitemShut {NoStop}%
\bibitem [{\citenamefont {Zhang}\ and\ \citenamefont {van Duin}(2017)}]{Zhang2017_2ndgen}%
  \BibitemOpen
  \bibfield  {author} {\bibinfo {author} {\bibfnamefont {W.}~\bibnamefont {Zhang}}\ and\ \bibinfo {author} {\bibfnamefont {A.~C.~T.}\ \bibnamefont {van Duin}},\ }\bibfield  {title} {\bibinfo {title} {Second-generation reaxff water force field: Improvements in the description of water density and oh-anion diffusion},\ }\href {https://doi.org/10.1021/acs.jpcb.7b02548} {\bibfield  {journal} {\bibinfo  {journal} {J. Phys. Chem. B}\ }\textbf {\bibinfo {volume} {121}},\ \bibinfo {pages} {6021} (\bibinfo {year} {2017})}\BibitemShut {NoStop}%
\bibitem [{\citenamefont {Zhang}\ and\ \citenamefont {van Duin}(2018)}]{Zhang2018_CHONweak}%
  \BibitemOpen
  \bibfield  {author} {\bibinfo {author} {\bibfnamefont {W.}~\bibnamefont {Zhang}}\ and\ \bibinfo {author} {\bibfnamefont {A.~C.~T.}\ \bibnamefont {van Duin}},\ }\bibfield  {title} {\bibinfo {title} {Improvement of the reaxff description for functionalized hydrocarbon/water weak interactions in the condensed phase},\ }\href {https://doi.org/10.1021/acs.jpcb.8b01127} {\bibfield  {journal} {\bibinfo  {journal} {J. Phys. Chem. B}\ }\textbf {\bibinfo {volume} {122}},\ \bibinfo {pages} {4083–4092} (\bibinfo {year} {2018})}\BibitemShut {NoStop}%
\bibitem [{\citenamefont {Molinero}\ and\ \citenamefont {Moore}(2009)}]{Molinero2009_mW}%
  \BibitemOpen
  \bibfield  {author} {\bibinfo {author} {\bibfnamefont {V.}~\bibnamefont {Molinero}}\ and\ \bibinfo {author} {\bibfnamefont {E.~B.}\ \bibnamefont {Moore}},\ }\bibfield  {title} {\bibinfo {title} {Water modeled as an intermediate element between carbon and silicon},\ }\href {https://doi.org/10.1021/jp805227c} {\bibfield  {journal} {\bibinfo  {journal} {J. Phys. Chem. B}\ }\textbf {\bibinfo {volume} {113}},\ \bibinfo {pages} {4008} (\bibinfo {year} {2009})},\ \bibinfo {note} {pMID: 18956896},\ \Eprint {https://arxiv.org/abs/https://doi.org/10.1021/jp805227c} {https://doi.org/10.1021/jp805227c} \BibitemShut {NoStop}%
\bibitem [{\citenamefont {Stillinger}\ and\ \citenamefont {Weber}(1985)}]{Stillinger1985_SWpotential}%
  \BibitemOpen
  \bibfield  {author} {\bibinfo {author} {\bibfnamefont {F.~H.}\ \bibnamefont {Stillinger}}\ and\ \bibinfo {author} {\bibfnamefont {T.~A.}\ \bibnamefont {Weber}},\ }\bibfield  {title} {\bibinfo {title} {Computer simulation of local order in condensed phases of silicon},\ }\href {https://doi.org/10.1103/PhysRevB.31.5262} {\bibfield  {journal} {\bibinfo  {journal} {Phys. Rev. B}\ }\textbf {\bibinfo {volume} {31}},\ \bibinfo {pages} {5262} (\bibinfo {year} {1985})}\BibitemShut {NoStop}%
\bibitem [{\citenamefont {Allen}\ and\ \citenamefont {Tildesley}(1987)}]{allen-tildesley-87}%
  \BibitemOpen
  \bibfield  {author} {\bibinfo {author} {\bibfnamefont {M.}~\bibnamefont {Allen}}\ and\ \bibinfo {author} {\bibfnamefont {D.}~\bibnamefont {Tildesley}},\ }\href@noop {} {\emph {\bibinfo {title} {{Computer Simulation of Liquids}}}}\ (\bibinfo  {publisher} {Oxford: Clarendon Pr},\ \bibinfo {year} {1987})\BibitemShut {NoStop}%
\bibitem [{\citenamefont {Thompson}\ \emph {et~al.}(2022)\citenamefont {Thompson}, \citenamefont {Aktulga}, \citenamefont {Berger}, \citenamefont {Bolintineanu}, \citenamefont {Brown}, \citenamefont {Crozier}, \citenamefont {in~'t Veld}, \citenamefont {Kohlmeyer}, \citenamefont {Moore}, \citenamefont {Nguyen}, \citenamefont {Shan}, \citenamefont {Stevens}, \citenamefont {Tranchida}, \citenamefont {Trott},\ and\ \citenamefont {Plimpton}}]{LAMMPS}%
  \BibitemOpen
  \bibfield  {author} {\bibinfo {author} {\bibfnamefont {A.~P.}\ \bibnamefont {Thompson}}, \bibinfo {author} {\bibfnamefont {H.~M.}\ \bibnamefont {Aktulga}}, \bibinfo {author} {\bibfnamefont {R.}~\bibnamefont {Berger}}, \bibinfo {author} {\bibfnamefont {D.~S.}\ \bibnamefont {Bolintineanu}}, \bibinfo {author} {\bibfnamefont {W.~M.}\ \bibnamefont {Brown}}, \bibinfo {author} {\bibfnamefont {P.~S.}\ \bibnamefont {Crozier}}, \bibinfo {author} {\bibfnamefont {P.~J.}\ \bibnamefont {in~'t Veld}}, \bibinfo {author} {\bibfnamefont {A.}~\bibnamefont {Kohlmeyer}}, \bibinfo {author} {\bibfnamefont {S.~G.}\ \bibnamefont {Moore}}, \bibinfo {author} {\bibfnamefont {T.~D.}\ \bibnamefont {Nguyen}}, \bibinfo {author} {\bibfnamefont {R.}~\bibnamefont {Shan}}, \bibinfo {author} {\bibfnamefont {M.~J.}\ \bibnamefont {Stevens}}, \bibinfo {author} {\bibfnamefont {J.}~\bibnamefont {Tranchida}}, \bibinfo {author} {\bibfnamefont {C.}~\bibnamefont {Trott}},\ and\ \bibinfo {author} {\bibfnamefont {S.~J.}\ \bibnamefont {Plimpton}},\
  }\bibfield  {title} {\bibinfo {title} {{LAMMPS} - a flexible simulation tool for particle-based materials modeling at the atomic, meso, and continuum scales},\ }\href {https://doi.org/10.1016/j.cpc.2021.108171} {\bibfield  {journal} {\bibinfo  {journal} {Comp. Phys. Comm.}\ }\textbf {\bibinfo {volume} {271}},\ \bibinfo {pages} {108171} (\bibinfo {year} {2022})}\BibitemShut {NoStop}%
\bibitem [{\citenamefont {Gittus}\ and\ \citenamefont {Bresme}(2021)}]{Gittus2021_ReaxFFWater}%
  \BibitemOpen
  \bibfield  {author} {\bibinfo {author} {\bibfnamefont {O.~R.}\ \bibnamefont {Gittus}}\ and\ \bibinfo {author} {\bibfnamefont {F.}~\bibnamefont {Bresme}},\ }\bibfield  {title} {\bibinfo {title} {Thermophysical properties of water using reactive force fields},\ }\href {https://doi.org/10.1063/5.0057868} {\bibfield  {journal} {\bibinfo  {journal} {J. Chem. Phys.}\ }\textbf {\bibinfo {volume} {155}},\ \bibinfo {pages} {114501} (\bibinfo {year} {2021})},\ \Eprint {https://arxiv.org/abs/https://doi.org/10.1063/5.0057868} {https://doi.org/10.1063/5.0057868} \BibitemShut {NoStop}%
\bibitem [{\citenamefont {Chapman}\ and\ \citenamefont {Bresme}(2022)}]{Chapman2022_TPwater}%
  \BibitemOpen
  \bibfield  {author} {\bibinfo {author} {\bibfnamefont {A.}~\bibnamefont {Chapman}}\ and\ \bibinfo {author} {\bibfnamefont {F.}~\bibnamefont {Bresme}},\ }\bibfield  {title} {\bibinfo {title} {Polarisation of water under thermal fields: the effect of the molecular dipole and quadrupole moments},\ }\href {https://doi.org/10.1039/D2CP00756H} {\bibfield  {journal} {\bibinfo  {journal} {Phys. Chem. Chem. Phys.}\ }\textbf {\bibinfo {volume} {24}},\ \bibinfo {pages} {14924} (\bibinfo {year} {2022})}\BibitemShut {NoStop}%
\bibitem [{\citenamefont {Ohara}(1999)}]{Ohara1999}%
  \BibitemOpen
  \bibfield  {author} {\bibinfo {author} {\bibfnamefont {T.}~\bibnamefont {Ohara}},\ }\bibfield  {title} {\bibinfo {title} {Intermolecular energy transfer in liquid water and its contribution to heat conduction: A molecular dynamics study},\ }\href {https://doi.org/10.1063/1.480025} {\bibfield  {journal} {\bibinfo  {journal} {J. Chem. Phys.}\ }\textbf {\bibinfo {volume} {111}},\ \bibinfo {pages} {6492} (\bibinfo {year} {1999})}\BibitemShut {NoStop}%
\bibitem [{\citenamefont {Bedrov}\ and\ \citenamefont {Smith}(2000)}]{Bedrov2000}%
  \BibitemOpen
  \bibfield  {author} {\bibinfo {author} {\bibfnamefont {D.}~\bibnamefont {Bedrov}}\ and\ \bibinfo {author} {\bibfnamefont {G.~D.}\ \bibnamefont {Smith}},\ }\bibfield  {title} {\bibinfo {title} {Thermal conductivity of molecular fluids from molecular dynamics simulations: Application of a new imposed-flux method},\ }\href {https://doi.org/10.1063/1.1312309} {\bibfield  {journal} {\bibinfo  {journal} {J. Chem. Phys.}\ }\textbf {\bibinfo {volume} {113}},\ \bibinfo {pages} {8080} (\bibinfo {year} {2000})}\BibitemShut {NoStop}%
\bibitem [{\citenamefont {Bresme}(2001)}]{Bresme2001}%
  \BibitemOpen
  \bibfield  {author} {\bibinfo {author} {\bibfnamefont {F.}~\bibnamefont {Bresme}},\ }\bibfield  {title} {\bibinfo {title} {Equilibrium and nonequilibrium molecular-dynamics simulations of the central force model of water},\ }\href {https://doi.org/10.1063/1.1407288} {\bibfield  {journal} {\bibinfo  {journal} {J. Chem. Phys.}\ }\textbf {\bibinfo {volume} {115}},\ \bibinfo {pages} {7564} (\bibinfo {year} {2001})}\BibitemShut {NoStop}%
\bibitem [{\citenamefont {Zhang}\ \emph {et~al.}(2005)\citenamefont {Zhang}, \citenamefont {Lussetti}, \citenamefont {de~Souza},\ and\ \citenamefont {Müller-Plathe}}]{Zhang2005}%
  \BibitemOpen
  \bibfield  {author} {\bibinfo {author} {\bibfnamefont {M.}~\bibnamefont {Zhang}}, \bibinfo {author} {\bibfnamefont {E.}~\bibnamefont {Lussetti}}, \bibinfo {author} {\bibfnamefont {L.~E.~S.}\ \bibnamefont {de~Souza}},\ and\ \bibinfo {author} {\bibfnamefont {F.}~\bibnamefont {Müller-Plathe}},\ }\bibfield  {title} {\bibinfo {title} {Thermal conductivities of molecular liquids by reverse nonequilibrium molecular dynamics},\ }\href {https://doi.org/10.1021/jp0512255} {\bibfield  {journal} {\bibinfo  {journal} {J. Phys. Chem. B}\ }\textbf {\bibinfo {volume} {109}},\ \bibinfo {pages} {15060} (\bibinfo {year} {2005})}\BibitemShut {NoStop}%
\bibitem [{\citenamefont {Terao}\ and\ \citenamefont {Müller-Plathe}(2005)}]{Terao2005}%
  \BibitemOpen
  \bibfield  {author} {\bibinfo {author} {\bibfnamefont {T.}~\bibnamefont {Terao}}\ and\ \bibinfo {author} {\bibfnamefont {F.}~\bibnamefont {Müller-Plathe}},\ }\bibfield  {title} {\bibinfo {title} {A nonequilibrium molecular dynamics method for thermal conductivities based on thermal noise},\ }\href {https://doi.org/10.1063/1.1858858} {\bibfield  {journal} {\bibinfo  {journal} {J. Chem. Phys.}\ }\textbf {\bibinfo {volume} {122}},\ \bibinfo {pages} {081103} (\bibinfo {year} {2005})}\BibitemShut {NoStop}%
\bibitem [{\citenamefont {Evans}\ \emph {et~al.}(2007)\citenamefont {Evans}, \citenamefont {Fish},\ and\ \citenamefont {Keblinski}}]{William2007}%
  \BibitemOpen
  \bibfield  {author} {\bibinfo {author} {\bibfnamefont {W.}~\bibnamefont {Evans}}, \bibinfo {author} {\bibfnamefont {J.}~\bibnamefont {Fish}},\ and\ \bibinfo {author} {\bibfnamefont {P.}~\bibnamefont {Keblinski}},\ }\bibfield  {title} {\bibinfo {title} {Thermal conductivity of ordered molecular water},\ }\href {https://doi.org/10.1063/1.2723071} {\bibfield  {journal} {\bibinfo  {journal} {J. Chem. Phys.}\ }\textbf {\bibinfo {volume} {126}},\ \bibinfo {pages} {154504} (\bibinfo {year} {2007})}\BibitemShut {NoStop}%
\bibitem [{\citenamefont {Rosenbaum}\ \emph {et~al.}(2007)\citenamefont {Rosenbaum}, \citenamefont {English}, \citenamefont {Johnson}, \citenamefont {Shaw},\ and\ \citenamefont {Warzinski}}]{Rosenbaum2007}%
  \BibitemOpen
  \bibfield  {author} {\bibinfo {author} {\bibfnamefont {E.~J.}\ \bibnamefont {Rosenbaum}}, \bibinfo {author} {\bibfnamefont {N.~J.}\ \bibnamefont {English}}, \bibinfo {author} {\bibfnamefont {J.~K.}\ \bibnamefont {Johnson}}, \bibinfo {author} {\bibfnamefont {D.~W.}\ \bibnamefont {Shaw}},\ and\ \bibinfo {author} {\bibfnamefont {R.~P.}\ \bibnamefont {Warzinski}},\ }\bibfield  {title} {\bibinfo {title} {Thermal conductivity of methane hydrate from experiment and molecular simulation},\ }\href {https://doi.org/10.1021/jp074419o} {\bibfield  {journal} {\bibinfo  {journal} {J. Phys. Chem. B}\ }\textbf {\bibinfo {volume} {111}},\ \bibinfo {pages} {13194} (\bibinfo {year} {2007})}\BibitemShut {NoStop}%
\bibitem [{\citenamefont {Jiang}\ \emph {et~al.}(2008)\citenamefont {Jiang}, \citenamefont {Myshakin}, \citenamefont {Jordan},\ and\ \citenamefont {Warzinski}}]{Jiang2008}%
  \BibitemOpen
  \bibfield  {author} {\bibinfo {author} {\bibfnamefont {H.}~\bibnamefont {Jiang}}, \bibinfo {author} {\bibfnamefont {E.~M.}\ \bibnamefont {Myshakin}}, \bibinfo {author} {\bibfnamefont {K.~D.}\ \bibnamefont {Jordan}},\ and\ \bibinfo {author} {\bibfnamefont {R.~P.}\ \bibnamefont {Warzinski}},\ }\bibfield  {title} {\bibinfo {title} {Molecular dynamics simulations of the thermal conductivity of methane hydrate},\ }\href {https://doi.org/10.1021/jp802942v} {\bibfield  {journal} {\bibinfo  {journal} {J. Phys. Chem. B}\ }\textbf {\bibinfo {volume} {112}},\ \bibinfo {pages} {10207} (\bibinfo {year} {2008})}\BibitemShut {NoStop}%
\bibitem [{\citenamefont {Kuang}\ and\ \citenamefont {Gezelter}(2010)}]{Kuang2010}%
  \BibitemOpen
  \bibfield  {author} {\bibinfo {author} {\bibfnamefont {S.}~\bibnamefont {Kuang}}\ and\ \bibinfo {author} {\bibfnamefont {J.~D.}\ \bibnamefont {Gezelter}},\ }\bibfield  {title} {\bibinfo {title} {A gentler approach to rnemd: Nonisotropic velocity scaling for computing thermal conductivity and shear viscosity},\ }\href {https://doi.org/10.1063/1.3499947} {\bibfield  {journal} {\bibinfo  {journal} {J. Chem. Phys.}\ }\textbf {\bibinfo {volume} {133}},\ \bibinfo {pages} {164101} (\bibinfo {year} {2010})}\BibitemShut {NoStop}%
\bibitem [{\citenamefont {Muscatello}\ and\ \citenamefont {Bresme}(2011)}]{Muscatello2011_JCP}%
  \BibitemOpen
  \bibfield  {author} {\bibinfo {author} {\bibfnamefont {J.}~\bibnamefont {Muscatello}}\ and\ \bibinfo {author} {\bibfnamefont {F.}~\bibnamefont {Bresme}},\ }\bibfield  {title} {\bibinfo {title} {A comparison of coulombic interaction methods in non-equilibrium studies of heat transfer in water},\ }\href {https://doi.org/10.1063/1.3670965} {\bibfield  {journal} {\bibinfo  {journal} {J. Chem. Phys.}\ }\textbf {\bibinfo {volume} {135}},\ \bibinfo {pages} {234111} (\bibinfo {year} {2011})}\BibitemShut {NoStop}%
\bibitem [{\citenamefont {Muscatello}\ \emph {et~al.}(2011)\citenamefont {Muscatello}, \citenamefont {Römer}, \citenamefont {Sala},\ and\ \citenamefont {Bresme}}]{Muscatello2011_ThermalPolarizationPCCP}%
  \BibitemOpen
  \bibfield  {author} {\bibinfo {author} {\bibfnamefont {J.}~\bibnamefont {Muscatello}}, \bibinfo {author} {\bibfnamefont {F.}~\bibnamefont {Römer}}, \bibinfo {author} {\bibfnamefont {J.}~\bibnamefont {Sala}},\ and\ \bibinfo {author} {\bibfnamefont {F.}~\bibnamefont {Bresme}},\ }\bibfield  {title} {\bibinfo {title} {Water under temperature gradients: polarization effects and microscopic mechanisms of heat transfer},\ }\href {https://doi.org/10.1039/C1CP21895F} {\bibfield  {journal} {\bibinfo  {journal} {Phys. Chem. Chem. Phys.}\ }\textbf {\bibinfo {volume} {13}},\ \bibinfo {pages} {19970} (\bibinfo {year} {2011})}\BibitemShut {NoStop}%
\bibitem [{\citenamefont {Römer}\ \emph {et~al.}(2012)\citenamefont {Römer}, \citenamefont {Lervik},\ and\ \citenamefont {Bresme}}]{Romer2012_waterTC}%
  \BibitemOpen
  \bibfield  {author} {\bibinfo {author} {\bibfnamefont {F.}~\bibnamefont {Römer}}, \bibinfo {author} {\bibfnamefont {A.}~\bibnamefont {Lervik}},\ and\ \bibinfo {author} {\bibfnamefont {F.}~\bibnamefont {Bresme}},\ }\bibfield  {title} {\bibinfo {title} {Nonequilibrium molecular dynamics simulations of the thermal conductivity of water: A systematic investigation of the spc/e and tip4p/2005 models},\ }\href {https://doi.org/10.1063/1.4739855} {\bibfield  {journal} {\bibinfo  {journal} {J. Chem. Phys.}\ }\textbf {\bibinfo {volume} {137}},\ \bibinfo {pages} {074503} (\bibinfo {year} {2012})}\BibitemShut {NoStop}%
\bibitem [{\citenamefont {Mao}\ and\ \citenamefont {Zhang}(2012)}]{Mao2012}%
  \BibitemOpen
  \bibfield  {author} {\bibinfo {author} {\bibfnamefont {Y.}~\bibnamefont {Mao}}\ and\ \bibinfo {author} {\bibfnamefont {Y.}~\bibnamefont {Zhang}},\ }\bibfield  {title} {\bibinfo {title} {Thermal conductivity, shear viscosity and specific heat of rigid water models},\ }\href {https://doi.org/10.1016/j.cplett.2012.05.044} {\bibfield  {journal} {\bibinfo  {journal} {Chem. Phys. Lett.}\ }\textbf {\bibinfo {volume} {542}},\ \bibinfo {pages} {37} (\bibinfo {year} {2012})}\BibitemShut {NoStop}%
\bibitem [{\citenamefont {Sirk}\ \emph {et~al.}(2013)\citenamefont {Sirk}, \citenamefont {Moore},\ and\ \citenamefont {Brown}}]{Sirk2013}%
  \BibitemOpen
  \bibfield  {author} {\bibinfo {author} {\bibfnamefont {T.~W.}\ \bibnamefont {Sirk}}, \bibinfo {author} {\bibfnamefont {S.}~\bibnamefont {Moore}},\ and\ \bibinfo {author} {\bibfnamefont {E.~F.}\ \bibnamefont {Brown}},\ }\bibfield  {title} {\bibinfo {title} {Characteristics of thermal conductivity in classical water models},\ }\href {https://doi.org/10.1063/1.4789961} {\bibfield  {journal} {\bibinfo  {journal} {J. Chem. Phys.}\ }\textbf {\bibinfo {volume} {138}},\ \bibinfo {pages} {064505} (\bibinfo {year} {2013})}\BibitemShut {NoStop}%
\bibitem [{\citenamefont {Lee}(2014)}]{Lee2014}%
  \BibitemOpen
  \bibfield  {author} {\bibinfo {author} {\bibfnamefont {S.~H.}\ \bibnamefont {Lee}},\ }\bibfield  {title} {\bibinfo {title} {Temperature dependence of the thermal conductivity of water: a molecular dynamics simulation study using the spc/e model},\ }\href {https://doi.org/10.1080/00268976.2014.891769} {\bibfield  {journal} {\bibinfo  {journal} {Mol. Phys.}\ }\textbf {\bibinfo {volume} {112}},\ \bibinfo {pages} {2155} (\bibinfo {year} {2014})}\BibitemShut {NoStop}%
\bibitem [{\citenamefont {Lee}\ and\ \citenamefont {Kim}(2019)}]{Lee2019}%
  \BibitemOpen
  \bibfield  {author} {\bibinfo {author} {\bibfnamefont {S.~H.}\ \bibnamefont {Lee}}\ and\ \bibinfo {author} {\bibfnamefont {J.}~\bibnamefont {Kim}},\ }\bibfield  {title} {\bibinfo {title} {Transport properties of bulk water at 243–550 k: a comparative molecular dynamics simulation study using spc/e, tip4p, and tip4p/2005 water models},\ }\href {https://doi.org/10.1080/00268976.2018.1562123} {\bibfield  {journal} {\bibinfo  {journal} {Mol. Phys.}\ }\textbf {\bibinfo {volume} {117}},\ \bibinfo {pages} {1926} (\bibinfo {year} {2019})}\BibitemShut {NoStop}%
\bibitem [{\citenamefont {Bresme}\ \emph {et~al.}(2014)\citenamefont {Bresme}, \citenamefont {Biddle}, \citenamefont {Sengers},\ and\ \citenamefont {Anisimov}}]{Bresme2014}%
  \BibitemOpen
  \bibfield  {author} {\bibinfo {author} {\bibfnamefont {F.}~\bibnamefont {Bresme}}, \bibinfo {author} {\bibfnamefont {J.~W.}\ \bibnamefont {Biddle}}, \bibinfo {author} {\bibfnamefont {J.~V.}\ \bibnamefont {Sengers}},\ and\ \bibinfo {author} {\bibfnamefont {M.~A.}\ \bibnamefont {Anisimov}},\ }\bibfield  {title} {\bibinfo {title} {Communication: Minimum in the thermal conductivity of supercooled water: A computer simulation study},\ }\href {https://doi.org/10.1063/1.4873167} {\bibfield  {journal} {\bibinfo  {journal} {J. Chem. Phys.}\ }\textbf {\bibinfo {volume} {140}},\ \bibinfo {pages} {161104} (\bibinfo {year} {2014})}\BibitemShut {NoStop}%
\bibitem [{\citenamefont {English}\ and\ \citenamefont {Tse}(2014)}]{English2014}%
  \BibitemOpen
  \bibfield  {author} {\bibinfo {author} {\bibfnamefont {N.~J.}\ \bibnamefont {English}}\ and\ \bibinfo {author} {\bibfnamefont {J.~S.}\ \bibnamefont {Tse}},\ }\bibfield  {title} {\bibinfo {title} {Thermal conductivity of supercooled water: An equilibrium molecular dynamics exploration},\ }\href {https://doi.org/10.1021/jz5016179} {\bibfield  {journal} {\bibinfo  {journal} {J. Phys. Chem. Lett.}\ }\textbf {\bibinfo {volume} {5}},\ \bibinfo {pages} {3819} (\bibinfo {year} {2014})},\ \bibinfo {note} {pMID: 26278754},\ \Eprint {https://arxiv.org/abs/https://doi.org/10.1021/jz5016179} {https://doi.org/10.1021/jz5016179} \BibitemShut {NoStop}%
\bibitem [{\citenamefont {Xu}\ \emph {et~al.}(2023)\citenamefont {Xu}, \citenamefont {Hao}, \citenamefont {Liang}, \citenamefont {Ying}, \citenamefont {Xu}, \citenamefont {Wu},\ and\ \citenamefont {Fan}}]{Xu2023_waterMLP_TC}%
  \BibitemOpen
  \bibfield  {author} {\bibinfo {author} {\bibfnamefont {K.}~\bibnamefont {Xu}}, \bibinfo {author} {\bibfnamefont {Y.}~\bibnamefont {Hao}}, \bibinfo {author} {\bibfnamefont {T.}~\bibnamefont {Liang}}, \bibinfo {author} {\bibfnamefont {P.}~\bibnamefont {Ying}}, \bibinfo {author} {\bibfnamefont {J.}~\bibnamefont {Xu}}, \bibinfo {author} {\bibfnamefont {J.}~\bibnamefont {Wu}},\ and\ \bibinfo {author} {\bibfnamefont {Z.}~\bibnamefont {Fan}},\ }\bibfield  {title} {\bibinfo {title} {{Accurate prediction of heat conductivity of water by a neuroevolution potential}},\ }\href {https://doi.org/10.1063/5.0147039} {\bibfield  {journal} {\bibinfo  {journal} {J. Chem. Phys.}\ }\textbf {\bibinfo {volume} {158}},\ \bibinfo {pages} {204114} (\bibinfo {year} {2023})},\ \Eprint {https://arxiv.org/abs/https://pubs.aip.org/aip/jcp/article-pdf/doi/10.1063/5.0147039/18052943/204114\_1\_5.0147039.pdf} {https://pubs.aip.org/aip/jcp/article-pdf/doi/10.1063/5.0147039/18052943/204114\_1\_5.0147039.pdf} \BibitemShut {NoStop}%
\bibitem [{\citenamefont {Matsubara}\ \emph {et~al.}(2021)\citenamefont {Matsubara}, \citenamefont {Kikugawa},\ and\ \citenamefont {Ohara}}]{Matsubara2021_MolecHeatTransfer}%
  \BibitemOpen
  \bibfield  {author} {\bibinfo {author} {\bibfnamefont {H.}~\bibnamefont {Matsubara}}, \bibinfo {author} {\bibfnamefont {G.}~\bibnamefont {Kikugawa}},\ and\ \bibinfo {author} {\bibfnamefont {T.}~\bibnamefont {Ohara}},\ }\bibfield  {title} {\bibinfo {title} {Comparison of molecular heat transfer mechanisms between water and ammonia in the liquid states},\ }\href {https://doi.org/https://doi.org/10.1016/j.ijthermalsci.2020.106762} {\bibfield  {journal} {\bibinfo  {journal} {Int. J. Therm. Sci.}\ }\textbf {\bibinfo {volume} {161}},\ \bibinfo {pages} {106762} (\bibinfo {year} {2021})}\BibitemShut {NoStop}%
\bibitem [{\citenamefont {Barnard}\ and\ \citenamefont {Russo}(2002)}]{Barnard2002_SWcarbon}%
  \BibitemOpen
  \bibfield  {author} {\bibinfo {author} {\bibfnamefont {A.~S.}\ \bibnamefont {Barnard}}\ and\ \bibinfo {author} {\bibfnamefont {S.~P.}\ \bibnamefont {Russo}},\ }\bibfield  {title} {\bibinfo {title} {Development of an improved stillinger-weber potential for tetrahedral carbon using ab initio (hartree-fock and mp2) methods},\ }\href@noop {} {\bibfield  {journal} {\bibinfo  {journal} {Mol. Phys.}\ }\textbf {\bibinfo {volume} {100}},\ \bibinfo {pages} {1517} (\bibinfo {year} {2002})}\BibitemShut {NoStop}%
\bibitem [{\citenamefont {Skinner}\ \emph {et~al.}(2014)\citenamefont {Skinner}, \citenamefont {Benmore}, \citenamefont {Neuefeind},\ and\ \citenamefont {Parise}}]{Skinner2014}%
  \BibitemOpen
  \bibfield  {author} {\bibinfo {author} {\bibfnamefont {L.~B.}\ \bibnamefont {Skinner}}, \bibinfo {author} {\bibfnamefont {C.~J.}\ \bibnamefont {Benmore}}, \bibinfo {author} {\bibfnamefont {J.~C.}\ \bibnamefont {Neuefeind}},\ and\ \bibinfo {author} {\bibfnamefont {J.~B.}\ \bibnamefont {Parise}},\ }\bibfield  {title} {\bibinfo {title} {The structure of water around the compressibility minimum},\ }\href {https://doi.org/10.1063/1.4902412} {\bibfield  {journal} {\bibinfo  {journal} {J. Chem. Phys.}\ }\textbf {\bibinfo {volume} {141}},\ \bibinfo {pages} {214507} (\bibinfo {year} {2014})}\BibitemShut {NoStop}%
\bibitem [{\citenamefont {Huang}\ \emph {et~al.}(2011)\citenamefont {Huang}, \citenamefont {Wikfeldt}, \citenamefont {Nordlund}, \citenamefont {Bergmann}, \citenamefont {McQueen}, \citenamefont {Sellberg}, \citenamefont {Pettersson},\ and\ \citenamefont {Nilsson}}]{Huang2011}%
  \BibitemOpen
  \bibfield  {author} {\bibinfo {author} {\bibfnamefont {C.}~\bibnamefont {Huang}}, \bibinfo {author} {\bibfnamefont {K.~T.}\ \bibnamefont {Wikfeldt}}, \bibinfo {author} {\bibfnamefont {D.}~\bibnamefont {Nordlund}}, \bibinfo {author} {\bibfnamefont {U.}~\bibnamefont {Bergmann}}, \bibinfo {author} {\bibfnamefont {T.}~\bibnamefont {McQueen}}, \bibinfo {author} {\bibfnamefont {J.}~\bibnamefont {Sellberg}}, \bibinfo {author} {\bibfnamefont {L.~G.~M.}\ \bibnamefont {Pettersson}},\ and\ \bibinfo {author} {\bibfnamefont {A.}~\bibnamefont {Nilsson}},\ }\bibfield  {title} {\bibinfo {title} {Wide-angle x-ray diffraction and molecular dynamics study of medium-range order in ambient and hot water},\ }\href {https://doi.org/10.1039/C1CP22804H} {\bibfield  {journal} {\bibinfo  {journal} {Phys. Chem. Chem. Phys.}\ }\textbf {\bibinfo {volume} {13}},\ \bibinfo {pages} {19997} (\bibinfo {year} {2011})}\BibitemShut {NoStop}%
\bibitem [{\citenamefont {Schlesinger}\ \emph {et~al.}(2016)\citenamefont {Schlesinger}, \citenamefont {Wikfeldt}, \citenamefont {Skinner}, \citenamefont {Benmore}, \citenamefont {Nilsson},\ and\ \citenamefont {Pettersson}}]{Schlesinger2016}%
  \BibitemOpen
  \bibfield  {author} {\bibinfo {author} {\bibfnamefont {D.}~\bibnamefont {Schlesinger}}, \bibinfo {author} {\bibfnamefont {K.~T.}\ \bibnamefont {Wikfeldt}}, \bibinfo {author} {\bibfnamefont {L.~B.}\ \bibnamefont {Skinner}}, \bibinfo {author} {\bibfnamefont {C.~J.}\ \bibnamefont {Benmore}}, \bibinfo {author} {\bibfnamefont {A.}~\bibnamefont {Nilsson}},\ and\ \bibinfo {author} {\bibfnamefont {L.~G.~M.}\ \bibnamefont {Pettersson}},\ }\bibfield  {title} {\bibinfo {title} {The temperature dependence of intermediate range oxygen-oxygen correlations in liquid water},\ }\href {https://doi.org/10.1063/1.4961404} {\bibfield  {journal} {\bibinfo  {journal} {J. Chem. Phys.}\ }\textbf {\bibinfo {volume} {145}},\ \bibinfo {pages} {084503} (\bibinfo {year} {2016})}\BibitemShut {NoStop}%
\bibitem [{\citenamefont {Nilsson}\ and\ \citenamefont {Pettersson}(2015)}]{Nilsson2015}%
  \BibitemOpen
  \bibfield  {author} {\bibinfo {author} {\bibfnamefont {A.}~\bibnamefont {Nilsson}}\ and\ \bibinfo {author} {\bibfnamefont {L.~G.~M.}\ \bibnamefont {Pettersson}},\ }\bibfield  {title} {\bibinfo {title} {The structural origin of anomalous properties of liquid water},\ }\href {https://doi.org/10.1038/ncomms9998} {\bibfield  {journal} {\bibinfo  {journal} {Nat. Commun.}\ }\textbf {\bibinfo {volume} {6}},\ \bibinfo {pages} {8998} (\bibinfo {year} {2015})}\BibitemShut {NoStop}%
\bibitem [{\citenamefont {Russo}\ and\ \citenamefont {Tanaka}(2014)}]{Tanaka2014}%
  \BibitemOpen
  \bibfield  {author} {\bibinfo {author} {\bibfnamefont {J.}~\bibnamefont {Russo}}\ and\ \bibinfo {author} {\bibfnamefont {H.}~\bibnamefont {Tanaka}},\ }\bibfield  {title} {\bibinfo {title} {Understanding water’s anomalies with locally favoured structures},\ }\href {https://doi.org/10.1038/ncomms4556} {\bibfield  {journal} {\bibinfo  {journal} {Nat. Commun.}\ }\textbf {\bibinfo {volume} {5}},\ \bibinfo {pages} {3556} (\bibinfo {year} {2014})}\BibitemShut {NoStop}%
\bibitem [{\citenamefont {Vedamuthu}\ \emph {et~al.}(1995)\citenamefont {Vedamuthu}, \citenamefont {Singh},\ and\ \citenamefont {Robinson}}]{Vedamuthu1995}%
  \BibitemOpen
  \bibfield  {author} {\bibinfo {author} {\bibfnamefont {M.}~\bibnamefont {Vedamuthu}}, \bibinfo {author} {\bibfnamefont {S.}~\bibnamefont {Singh}},\ and\ \bibinfo {author} {\bibfnamefont {G.~W.}\ \bibnamefont {Robinson}},\ }\bibfield  {title} {\bibinfo {title} {Properties of liquid water. 4. the isothermal compressibility minimum near 50 .degree.c},\ }\href {https://doi.org/10.1021/j100022a047} {\bibfield  {journal} {\bibinfo  {journal} {J. Phys. Chem.}\ }\textbf {\bibinfo {volume} {99}},\ \bibinfo {pages} {9263} (\bibinfo {year} {1995})}\BibitemShut {NoStop}%
\end{thebibliography}%

\end{document}